# Relative Dielectric Constants And Selectivity Ratios In Open Ionic Channels

**Bob Eisenberg**
Department of Applied Mathematics
Illinois Institute of Technology
Chicago IL
USA

Department of Physiology and Biophysics
Rush University
Chicago IL
USA
bob.eisenberg@gmail.com

**Weishi Liu**
Department of Mathematics
University of Kansas
Lawrence, Kansas 66045
wsliu@ku.edu

# Abstract

We investigate the effects of the relative dielectric coefficient on ionic flows in open ion channels, using mathematical analysis of reasonably general Poisson-Nernst-Planck type models that can include the finite sizes of ions. The value of the relative dielectric coefficient is of course a crucial parameter for ionic behavior in general. Using the powerful theory of singularly perturbed problems in applied mathematics, we show that some properties of open channels are quite insensitive to variation in the relative dielectric coefficient, thereby explaining such effects seen unexpectedly in simulations. The ratio between the total number of one ion species and that of another ion species, and the ratio between the flux of one ion species and that of another ion species do not depend significantly on the relative dielectric coefficient.

## 1    Introduction

For open channels, permeation and selectivity are among the most important properties of channel functions ([16, 17, 18, 21, 22, 31, 34, 36, 37, 38, 75, 76, 85, 91, 96]). These macroscopic properties are outputs of nonlinear interactions of these parameters, such as channel structures (channel shape and spatial distribution of permanent charges), boundary concentrations, electric potential differences, diffusion coefficients, dielectric properties, ionic sizes, and many others, through microscopic laws of electrodiffusion and chemical interactions (see, [3, 4, 5, 6, 10, 16, 35, 46, 47, 49, 51, 65, 73, 83, 86, 88, 90, 92, 93], etc.).

To understand these properties, the atomic structure of the channel protein is critically important (see, e.g., [11, 12, 13, 16, 28, 37, 92, 104, 106, 105]). It determines qualitatively mechanisms of selectivity and permeation, and provides basis for high resolutions of integrating correlations in continuum models. This approach with structural features does not though deal quantitatively with the experimental measurements of current at all as can be seen by the absence of predicted graphs of currents or contents vs. voltage or concentration. The structures being discussed are determined from crystals, with somewhat different locations of atoms and forces (or they would not crystallize), in solutions remote from the physiological solutions in which the channels function, and often at temperatures around 100K. Qualitative discussion is of great importance but it must not be confused with attempts at quantitative models.

The quantitative treatment of selectivity and permeation differs from the usual qualitative treatment in several important respects. The quantitative approach based on continuum models treats selectivity and permeation as expressions of the current flow through a channel under a variety of conditions, namely different membrane potentials, different concentrations and compositions of ionic solutions, and different mutations. The quantities from computations/analyses of models can be compared directly with experimental measurements of current. The quantitative models are dramatically reduced and oversimplified to be sure, but they are precise. Such is the nature of most physical models of condensed phases. The utility of the models is determined not by discussion of their oversimplifications but by their ability to fit wide ranges of data with single sets of parameter, and to design new systems and predict their properties. Physicists tend to emphasize the importance of fitting data; structural biologists tend to emphasize the importance of structural detail. Both approaches are needed and such complement and compliment each other, in our view.

Poisson-Nernst-Planck models were introduced to deal with the correlations produced by the mean electric field. Previous models assumed potentials that were constant as charges and structures changed. The goal of PNP was to deal with the change in potential profiles enforced by the Poisson part of the Maxwell equations. PNP was then refined to describe correlations of other types with more accuracy [4, 24, 25, 26, 27, 36, 41, 42, 43, 44, 49, 50, 51, 58, 59, 60, 61, 62, 63, 64, 65, 72, 77, 78, 79, 80, 86, 97, 98, 103] with considerable success for calcium channels (particularly the ryanodine receptor) and good (but incomplete) success for the DEKA sodium channel.

The essential point here is that models of the PNP type we have studied capture correlations well enough to explain the main experimental data from three major types of channels cited above (see [3] for more discussions). Ionic flow through biological channels exhibits extremely rich dynamics ([19, 20, 37, 38, 96], etc.). The flow is multi-scale (in both space and time) and the general behavior depends on many parameters such as channel structures (channel shape and spatial distribution of permanent charges), boundary concentrations, electric potential differences, diffusion coefficients, dielectric properties, ionic sizes, and many others ([3, 4, 17, 18, 21, 22]). Some of the parameters affect each other, for example, the dielectric coefficients and diffusion coefficients are by no means constants and they vary as the environment changes. It is a great challenge to understand the full behavior of ionic flows. On the other hand, not all the parameters play equal role for every biologically interesting quantity. For example, the *sign or direction* of the flux of an ion species depends only on the difference of its electrochemical potentials at the boundary points of the channel ([27]); in particular, the *sign* of the flux **does not** depend on the permanent charge (either its spatial occupancy and its density) and specifics of the dielectric coefficients; of course, the *magnitude* of the flux **does** depend on the permanent charge and specifics of dielectric coefficients as well as other physical parameters of the biological system. Another example concerns ion channels without permanent charge: the shape does not play a significant role in the *magnitude* of individual fluxes – an average property of the channel shape is the main determinant in this case ([48]).

This paper is not meant to produce complete models of any channel. Rather, motivated by the work in [3], we will examine a particular property of selectivity in this paper with the quantitative approach based on Poisson-Nernst-Planck (PNP) type models.

In this work, we are interested in the effects of the relative dielectric coefficient on ionic flow properties. The value of the relative dielectric coefficient is of course a crucial parameter for ionic behaviors in general. From the modeling point of view, the traditional Poisson-Boltzmann system with constant dielectric coefficients as in the Debye-Hückel theory has been modified with concentration-dependent dielectric coefficients and ion-water interaction that show significant improvements for the qualitative property of the mean activity coefficient (see, for example, [99, 100]).

We will examine some properties of ionic movements and show they are not sensitive to changes in the relative dielectric coefficient. For example, if we consider the ionic flow through a biological channel (for example, sodium channel in particular the DEKA aspartate-glutamate-lysine-alanine channel responsible for the rising phase of nerve action potentials) involving the mixture of $Na^+Cl^-$ and $K^+Cl^-$, the total number $\#(Na)$ of $Na^+$ ions and the total number $\#$

(K) of K$^+$ ions in the channel could depend on the value of the relative dielectric coefficient significantly, but the ratio # (Na)/ # K) between these two numbers does not depend significantly on the value of the relative dielectric coefficient under reasonable conditions. The latter has been observed in extensive Monte Carlo simulations (to the surprise of the authors, we are told, see Fig. 8-10 in [3]). Applying the powerful singular perturbation theory to Poisson-Nernst-Planck (PNP) type models that include finite sizes of ions, we provide analytical justifications **when the ratio between fluxes of different ion species and the ratio between total numbers of particles of different ion species of the ionic mixture do not depend significantly on the relative dielectric coefficient, and when the ratios might depend significantly on the relative dielectric coefficient.**

We emphasize that the dependences of the ratios on other physical parameters such as radius of the channel, protein structure, etc. are important. In the future, we will examine these dependences which require much more detailed analysis of PNP type models than that used in this paper for the present purpose.

We stress that our PNP type models can include finite sizes of ions and allow general distributions of permanent charges of the channel. *The key condition for our results is that the standard dimensionless parameter $\varepsilon$ defined in (4) is small*. This dimensionless parameter $\varepsilon$ is a version of the Debye length divided by the length $l$ of the channel that reflects a combined effect of the relative dielectric coefficient $\varepsilon_r$, the dielectric constant $\varepsilon_0$ of vacuum, and the characteristic concentration $C_0$ of the ionic mixture. The parameter $\varepsilon$ is small mostly because of the length scale of the channel and the characteristic concentration of the ionic mixture. For general electrolytes problems, the value of $\varepsilon$ could vary several orders of magnitudes. Our results *do not apply* when the value of $\varepsilon$ is large or moderate. For ion channel problems, the value of $\varepsilon$ would be small when the characteristic concentration $C_0$ and/or the length $l$ of channel are large. We estimate $\varepsilon$ to be about $10^{-2}$ or smaller for a few biological situations (see the end of Section 3.1 and Remark 3.1).

We remark that when the value of the relative dielectric coefficient is varied, we assume other properties such as diffusion coefficients, permanent charge distribution and shape of the channel remain more or less unchanged. The flexibility of these parameters would require more extensive analysis of PNP type models with possible modifications of the models. We will examine these important issues in future work.

# 2   Poisson-Nernst-Planck models

## 2.1   Three-dimensional model

For an ionic mixture with $n$ ion species, PNP reads

$$\nabla \cdot (\varepsilon_r(r)\varepsilon_0 \nabla \Phi) = -e(\sum_{s=1}^{n} z_s C_s + Q(r)),$$

$$\nabla \cdot J_k = 0, \quad -J_k = \frac{1}{k_B T} D_k(r) C_k \nabla \mu_k, \quad k = 1, 2, \cdots, n \tag{2.1}$$

where $r \in \Omega$. $\Omega$ is a three-dimensional cylindrical-like domain representing the channel, $Q(r)$ is the permanent charge density, $\varepsilon_r(r)$ is the relative dielectric coefficient, $\varepsilon_0$ is the vacuum permittivity, $e$ is the elementary charge, $k_B$ is the Boltzmann constant, $T$ is the absolute temperature; $\Phi$ is the electric potential, and, for the $k$ th ion species, $C_k$ is the concentration, $z_k$ is the valence (the number of charges per particle), $\mu_k$ is the electrochemical potential depending on the concentrations $\{C_j\}$ and finite sizes of all ion species, $J_k$ is the flux density, and $D_k(r)$ is the diffusion coefficient.

### 2.2 Quasi-one-dimensional PNP models

Reduction of three-dimensional PNP systems (1) to quasi-one-dimensional models was first proposed in [82] using the fact that ion channels have narrow cross-sections relative to their lengths and was rigorously justified in [68] for special cases. A quasi-one-dimensional PNP model is

$$\frac{1}{A(X)}\frac{d}{dX}\left(\varepsilon_r(X)\varepsilon_0 A(X)\frac{d}{dX}\Phi\right) = -e(\sum_{s=1}^{n}z_s C_s + Q(X)),$$

$$\frac{d}{dX}J_k = 0, \quad -J_k = \frac{1}{k_B T}D_k(X)A(X)C_k\frac{d}{dX}\mu_k, \quad k = 1,2,\cdots,n \tag{2.2}$$

where $X \in [0,l]$ is the coordinate along the axis of the channel, $A(X)$ is the area of cross-section of the channel over the location $X$. Equipped with system (2), we impose the following boundary conditions (see, [26] for a reasoning), for $k = 1,2,\cdots,n$,

$$\Phi(0) = V, \ C_k(0) = L_k > 0; \quad \Phi(l) = 0, \ _k(l) = R_k > 0. \tag{2.3}$$

## 2.3 Rescaling of the quasi-one-dimensional PNP model

The following rescaling (see [29]) or its variations have been widely used for convenience of mathematical analysis. To do so, we first introduce several quantities.

Let $C_0$ be a characteristic concentration of the ion solution; for example, one can take $C_0$ to be the maximal value of the boundary concentrations and the absolute value of the permanent charge:

$$C_0 = \max\{L_1, L_2, \ldots, L_n, R_1, R_2, \ldots, R_n, \sup_{X \in [0,l]} |Q(X)|\},$$

where $\sup_{X \in [0,l]}|Q(X)|$ is the least upper bound of the function $|Q(X)|$ for $X \in [0,l]$. Also, let $D_0$ be the maximal value of all diffusion coefficients:

$$D_0 = \max\{\sup_{X \in [0,l]}D_1(X), \sup_{X \in [0,l]}D_2(X), \ldots, \sup_{X \in [0,l]}D_n(X)\},$$

and let $\bar{\varepsilon}_r = \sup_{X \in [0,l]}\varepsilon_r(X)$.

Using these quantities, we now make a dimensionless re-scaling of the variables in system (2) as follows.

$$\varepsilon^2 = \frac{\overline{\varepsilon}_r \varepsilon_0 k_B T}{e^2 l^2 C_0}, \quad x = \frac{X}{l}, \quad h(x) = \frac{A(X)}{l^2},$$

$$\hat{\varepsilon}_r(x) = \frac{\varepsilon_r(X)}{\overline{\varepsilon}_r}, \quad D_k(x) = \frac{D_k(X)}{D_0}, \quad Q(x) = \frac{Q(X)}{C_0}, \tag{2.4}$$

$$\phi(x) = \frac{e}{k_B T} \Phi(X), \quad c_k(x) = \frac{C_k(X)}{C_0}, \quad J_k = \frac{J_k}{l C_0 D_0}.$$

In terms of the new variables, the boundary value problem (BVP) associated to equation (2) and the boundary conditions (3) becomes

$$\frac{\varepsilon^2}{h(x)} \frac{d}{dx}\left(\hat{\varepsilon}_r(x) h(x) \frac{d\phi}{dx}\right) = -\sum_{s=1}^{n} z_s c_s - Q(x),$$

$$\frac{dJ_k}{dx} = 0, \quad \frac{1}{k_B T} D_k(x) h(x) c_k \frac{d\mu_k}{dx} = -J_k, \tag{2.5}$$

with boundary conditions at $x = 0$ and $x = 1$

$$\phi(0) = \quad V_0 := \frac{e}{k_B T} V, c_k(0) = l_k := \frac{L_k}{C_0};$$

$$\phi(1) = \quad 0, c_k(1) = r_k := \frac{R_k}{C_0}. \tag{2.6}$$

We comment that the electrochemical potential $\mu_k$ can include excess components $\mu_k^{ex}$ (beyond the ideal component $\mu_k^{id}$) for the ion size effect. For example, one can include any excess potential of the form

$$\frac{1}{k_B T} \mu_k^{ex} = \sum_{j=1}^{n} g_{kj} c_j$$

with $g_{jk}$'s depending on ion sizes (see, e.g., [41, 58, 59]), or some specific approximation models in [32, 33].

We also want to emphasize that our results concern a particular type of properties, that is, how the ratio of between fluxes of different ion species and the ratio between total numbers of particles of different ion species of the ionic mixture depend on variation of relative dielectric coefficient. We do not consider the effects of other quantities in this work. Our conclusion will be based on some properties, not the specific formulas which are not available in general, of the solutions of BVP (5) and (6).

# 3    Relevant properties of the singularly perturbed BVP and its consequences

A mathematical consequence of the key assumption of small $\varepsilon$ is that the BVP can be treated as a *singularly perturbed problem*. A general geometric framework for analyzing singularly perturbed BVP of PNP type systems (5), (6) has been developed in [26, 48, 66, 67, 69] for classical PNP systems and in [47, 57] for PNP systems with finite ion size. We give a brief discussion about the framework and refer readers to above mentioned references for details.

Due to the smallness of the parameter $\varepsilon$, the BVP (5) and (6) has two distinct scales: *fast scale* for boundary/transition layers and *slow scale* for bulk behaviors. Accordingly, the system (5) can be split into two complementary subsystems: *the fast subsystem* and *the slow subsystem*. These subsystems, in the $\varepsilon \to 0$ limits, become lower dimensional systems with the sum of their dimensions being that of the full system. Most importantly, these limiting lower dimensional systems correspond to "ideal" situations of physical problems and have a better chance to be more or less completely understood. This is the case for classical PNP systems ([26, 48, 66, 67, 69]) and for PNP systems with finite ion sizes ([47, 57]). Furthermore, the geometric singular perturbation theory provides a way that allows one to piece together the limiting information from the subsystems to the full system for $\varepsilon > 0$ small.

The essential result of the geometric singular perturbation theory for the singularly perturbed BVP (5) and (6) is (see, e.g., [26, 66, 67]), for small $\varepsilon > 0$, if

$$(\phi(x;\varepsilon), c_1(x;\varepsilon), c_2(x;\varepsilon), \ldots, c_n(x;\varepsilon), J_1(\varepsilon), J_2(\varepsilon), \ldots, J_n(\varepsilon))$$

is the solution, then, for each $1 \le k \le n$, one can express $c_k(x;\varepsilon)$ as

$$c_k(x;\varepsilon) = c_{k0}(x) + \varepsilon c_{k1}(x;\varepsilon), \tag{3.1}$$

where $c_{k0}(x)$, the zeroth order in $\varepsilon$ approximation, is independent of $\varepsilon$ and is continuous at all $x$ *except at those points where the permanent charge $Q(x)$ jumps* (discontinuity points of $Q(x)$). Similarly, for each $1 \le k \le n$,

$$J_k(\varepsilon) = J_{k0} + \varepsilon J_{k1}(\varepsilon), \tag{3.2}$$

where $J_{k0}$ is a constant independent of $\varepsilon$.

While quantities $c_{k0}(x)$ and $J_{k0}$ in (1) and (2) depend on parameters such as boundary conditions and permanent charge distributions, they do not depend on $\varepsilon$. The terms $\varepsilon c_{k1}(x;\varepsilon)$ and $\varepsilon J_{k1}(\varepsilon)$ depend on $\varepsilon$, in addition to other parameters, but in higher order terms in $\varepsilon$.

We comment that, in general, for example, when permanent charges and/or ion sizes are present, analytical formulas for the zeroth order terms $c_{k0}(x)$ and $J_{k0}$ are not available. In special cases, such as, zero or small permanent charges and/or ion sizes, $c_{k0}(x)$ and $J_{k0}$ depend on permanent charges and ion sizes and the leading orders of $c_{k0}(x)$ and $J_{k0}$ in permanent charges and ion sizes have explicit analytic formulas ([48, 57]).

In Figure 1 (next page), the above property for the concentrations in (1) is illustrated by numerical simulations using Matlab-bvp4c. (Thanks to Liwei Zhang for providing the numeric results.) The setup is for $n = 2$, $z_1 = 1$, $z_2 = -1$, $Q(x) = 0$ for $x \in (0,1/3) \cup (2/3,1)$ and $Q(x) = Q = 4$ for $x \in (1/3, 2/3)$ (so that the permanent charge $Q(x)$ has a jump discontinuity at $x = 1/3$ and at $x = 2/3$), and $\varepsilon = 0.05$ for the profiles in the left panel and $\varepsilon = 0.01$ for the profiles in the right panel. The boundary conditions in (6) are taken to be $V_0 = -0.5$, $l_1 = l_2 = L = 2$, $r_1 = r_2 = R = 3$

The figure shows numerical solutions of BVP associated to the PNP system (5) with ideal electrochemical potential and the boundary condition (6), where $\phi(x;\varepsilon)$ is the starred curve, $c_1(x;\varepsilon)$ is the solid curve and $c_2(x;\varepsilon))$ is the dashed curve. The profiles of $\phi(x;\varepsilon)$, $c_1(x;\varepsilon)$ and $c_2(x;\varepsilon)$ have layers around the discontinuity points $x=1/3$ and $x=2/3$ of $Q(x)$. As $\varepsilon$ changes from $0.05$ in the left panel to a smaller value $0.01$ in the right panel, the layers become sharp. In the $\varepsilon \to 0$ limits, the orbits of $c_1(x;\varepsilon)$ and $c_2(x;\varepsilon)$ in (1) would limit to those of $c_{10}(x)$ and $c_{20}(x)$ with vertical line segments as limiting layers; in particular, $c_{10}(x)$ and $c_{20}(x)$ would satisfy the electroneutrality condition $z_1 c_{10}(x) + z_2 c_{20}(x) + Q(x) = 0$ at every $x$ except $x=1/3$ and $x=2/3$ where $Q(x)$ has a jump discontinuity, and $c_{10}(x)$ and $c_{20}(x)$ have jump discontinuities at, and only at, $x=1/3$ and $x=2/3$.

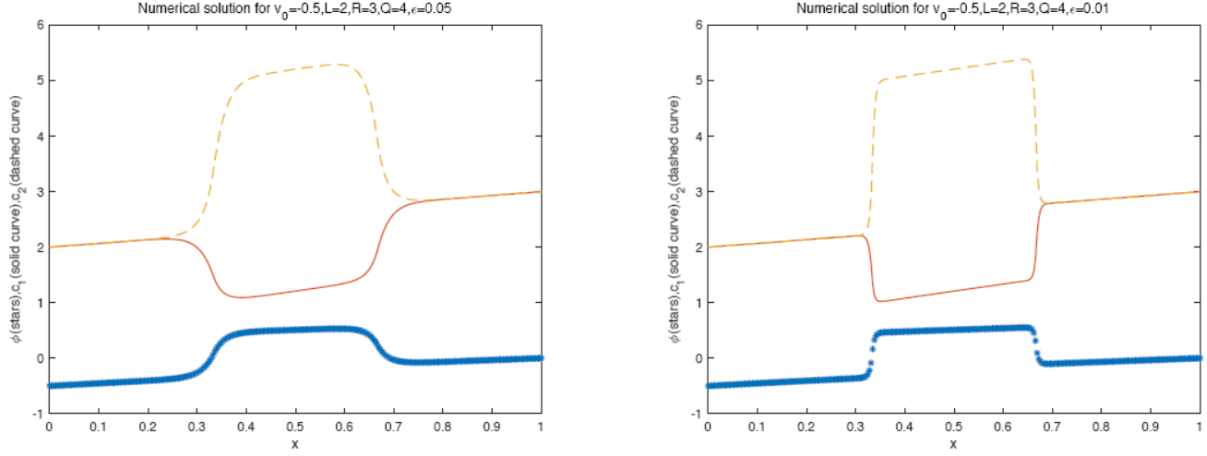

Figure 1: *A numerical solution of BVP (5) and (6) with* $n=2$, $z_1=1$, $z_2=-1$ $\varepsilon = 0.05$ *(left panel) and* $\varepsilon = 0.01$ *(right panel),* $\hat{\varepsilon}_r(x) = 1$, $Q(x) = 0$ *for* $x \in (0,1/3) \cup (2/3,1)$ *and* $Q(x) = Q = 4$ *for* $x \in (1/3,2/3)$, $V_0 = -0.5$, $l_1 = l_2 = L = 2$ *and* $r_1 = r_2 = R = 3$: $\phi(x;\varepsilon)$ *is the starred curve,* $c_1(x;\varepsilon)$ *is the solid curve, and* $c_2(x;\varepsilon)$ *is the dashed curve. Two values of* $\varepsilon$ *are used to illustrate the trend of sharpness of the layers as* $\varepsilon$ *changes from* $0.05$ *for the left panel to a smaller value* $0.01$ *for the right panel.*

Note that, in (1) and (2), the quantities $c_k(x;\varepsilon)$ and $J_k(\varepsilon)$ are scaled quantities in (4). In terms of the original unscaled quantities, we have

$$
\begin{aligned}
C_k(X;\varepsilon) &= C_0 c_k(x;\varepsilon) = C_0 c_{k0}(x) + \varepsilon C_0 c_{k1}(x), \\
J_k(\varepsilon) &= l C_0 D_0 J_k(\varepsilon) = l C_0 D_0 J_{k0} + \varepsilon l C_0 D_0 J_{k1}(\varepsilon).
\end{aligned}
\tag{3.3}
$$

Concerning $C_k$, it is noted from (4) that, when $C_0$ is large,

$$
\varepsilon = \sqrt{\frac{\bar{\varepsilon}_r \varepsilon_0 k_B T}{e^2 l^2 C_0}}
$$

is small, but

$$\varepsilon C_0 = \sqrt{\frac{\bar{\varepsilon}_r \varepsilon_0 k_B T C_0}{e^2 l^2}}$$

is **not** small in general (although, relative to $C_0$, it is small). Therefore, ignoring the second term $\varepsilon C_0 c_{k1}(x;\varepsilon)$ in (3) for the value of $C_k(x;\varepsilon)$ is tricky. Similar remark applies to the flux $J_k(\varepsilon)$.

### 3.1   Ratio of total numbers of different ion species

We now apply the results in (3) to the ratio of total numbers of two different ion species. For definiteness, we take $Na^+$ and $K^+$ ions.

Let $c_1(x;\varepsilon)$ and $c_2(x;\varepsilon)$ be the concentrations of $Na^+$ and $K^+$ in terms of the scaled variables, respectively. Then, the total numbers of $Na^+$ and $K^+$ ions in the channel are, respectively,

$$N_1(\varepsilon) = \int_0^l C_1(X;\varepsilon)dX \quad and \quad N_2(\varepsilon) = \int_0^l C_2(X;\varepsilon)dX.$$

It follows from $X = lx$ and $C_k(X;\varepsilon) = C_0 c_k(x;\varepsilon)$ in (4) that

$$\begin{aligned} N_1(\varepsilon) &= \int_0^l C_1(X;\varepsilon)dX = \int_0^1 l C_0 c_1(x;\varepsilon)dx, \\ N_2(\varepsilon) &= \int_0^l C_2(X;\varepsilon)dX = \int_0^1 l C_0 c_2(x;\varepsilon)dx. \end{aligned}$$

(3.4)

From extensive Monte Carlo simulations for DEKA Na channel in [3], it shows (Fig. 8-10 in [3]) that the ratio $N_1/N_2$ decreases as the channel diameter increases BUT the ratio does not depend on $\bar{\varepsilon}_r$ significantly, say, as $\bar{\varepsilon}_r$ changes from $1$ to $80$. We now provide an explanation to the latter from formulas (1) and (3).

It follows from (1) or (3) that

$$\begin{aligned} N_1(\varepsilon) &= l C_0 \int_0^1 c_{10}(x)dx + \varepsilon l C_0 \int_0^1 c_{11}(x;\varepsilon)dx, \\ N_2(\varepsilon) &= l C_0 \int_0^1 c_{20}(x)dx + \varepsilon l C_0 \int_0^1 c_{21}(x;\varepsilon)dx, \end{aligned}$$

(3.5)

and hence, the ratio between the total number of $Na^+$ and that of $K^+$ is

$$\frac{N_1(\varepsilon)}{N_2(\varepsilon)} = \frac{l C_0 \int_0^1 c_{10}(x)dx + \varepsilon l C_0 \int_0^1 c_{11}(x;\varepsilon)dx}{l C_0 \int_0^1 c_{20}(x)dx + \varepsilon l C_0 \int_0^1 c_{21}(x;\varepsilon)dx}$$

$$= \frac{\int_0^1 c_{10}(x)dx + \varepsilon \int_0^1 c_{11}(x;\varepsilon)dx}{\int_0^1 c_{20}(x)dx + \varepsilon \int_0^1 c_{21}(x;\varepsilon)dx}$$

(3.6)

$$= \frac{\int_0^1 c_{10}(x)dx}{\int_0^1 c_{20}(x)dx} + \varepsilon \frac{\int_0^1 c_{11}dx \int_0^1 c_{20}dx - \int_0^1 c_{10}dx \int_0^1 c_{21}dx}{(\int_0^1 c_{20}(x)dx)^2} + O(\varepsilon^2).$$

Note that the scaled variables $c_{k0}$ and $c_{k1}$ are of order $O(1)$ quantities. If the denominators in (6) are not small, that is,

$$\int_0^1 c_{20}(x)dx > 0 \text{ is not very small relative to } \varepsilon, \tag{3.7}$$

then the coefficient of the $\varepsilon$ order term is of order $O(1)$. In particular,

$$\frac{N_1(\varepsilon)}{N_2(\varepsilon)} \approx \frac{\int_0^1 c_{10}(x)dx}{\int_0^1 c_{20}(x)dx} \tag{3.8}$$

with the right-hand side independent of $\varepsilon$ and with an error bound in the order $O(\varepsilon)$. Note that, in the derivation (6), the large quantity $C_0$ is dropped when ratio $N_1(\varepsilon)/N_2(\varepsilon)$ is considered.

We now use the following specific values of the quantities involved to estimate the value of $\varepsilon$ in (4) for an assessment of the error bound.

- The length between the two baths containing the channel is $l(nm) = l \times 10^{-9} (m)$;
- $e \approx 1.60 \times 10^{-19}$ (C), $\quad \varepsilon_0 \approx 8.85 \times 10^{-12}$ (F·m$^{-1}$);
- $k_B \approx 1.38 \times 10^{-23}$ (JK$^{-1}$), $\quad T \approx 273$ (K).

Then, from (4),

$$\varepsilon^2 = \frac{k_B T \bar{\varepsilon}_r \varepsilon_0}{l^2 \times 10^{-18} \times C_0 \times N_A \times 10^3 e^2} \approx \bar{\varepsilon}_r \frac{2.16}{l^2 C_0} \times 10^{-3}.$$

If we take $l = 2.5$ (the length of the channel would be $2.5$ (nm)) and $C_0 = 10$ (the maximum of concentrations including the concentration of the permanent charge would be $10$ (M)), then

$$\varepsilon \approx \sqrt{\bar{\varepsilon}_r} \times 5.84 \times 10^{-3} \tag{3.9}$$

As relative dielectric coefficient $\bar{\varepsilon}_r$ changes from $1$ to $80$, the value of $\varepsilon$ changes by a factor of $\sqrt{80} \approx 9$. If the length $l$ of the channel is $2.5$ (nm) and the maximum of the concentration $C_0$ of the ionic solution is $10$ (M), then $\varepsilon$ is of order $10^{-3}$ (see (9)) so a factor of $9$ in the variation of $\bar{\varepsilon}_r$ makes about $5\%$ difference in the approximation (8) and is not significant.

**Remark 3.1** *Our quasi-one-dimensional model of a channel setup includes access regions and baths, following [68, 82]. The variable diameter (equivalently, the variable cross sectional area* $A(X)$ in (2)) of the 'channel' in this setup includes a region with the diameter of the actual pore of the channel protein, including the taper often seen in structures as they join the extracellular space. In our setup, the diameter of the 'channel' is dramatically increased outside the channel protein to mimic (crudely but adequately for most purposes) the huge increase in the cross sectional area available for current flow, as ions leave the channel protein and enter the extracellular space with its macroscopic cross sectional area. What is important is that resistance to electric current flow and effective 'resistance' to diffusional flow is millions of times less in the bath region than inside the channel protein. Thus, in this treatment the quantity $l$ is the length between the two electrodes where boundary conditions (3) are imposed. The channel would correspond to a subinterval of $[0, l]$, or in the scaled variable $x$, to a subinterval, say $[a, b]$, of the interval $[0, 1]$. For this setting, $c_1(x; \varepsilon)$ and $c_2(x; \varepsilon)$ would be different from those in the previous setting. The total numbers of $Na^+$ and $K^+$ ions in the channel would be, respectively,

$$N_1(\varepsilon) = \int_a^b l C_0 c_1(x; \varepsilon) dx \quad and \quad N_2(\varepsilon) = \int_a^b l C_0 c_2(x; \varepsilon) dx.$$

Following the lines of derivation in (6), the ratio between the total number of $Na^+$ and that of $K^+$ would be

$$\frac{N_1(\varepsilon)}{N_2(\varepsilon)} = \frac{\int_a^b c_{10}(x) dx}{\int_a^b c_{20}(x) dx} + \varepsilon \frac{\int_a^b c_{11} dx \int_a^b c_{20} dx - \int_a^b c_{10} dx \int_a^b c_{21} dx}{(\int_a^b c_{20}(x) dx)^2} + O(\varepsilon^2).$$

Hence,

$$\frac{N_1(\varepsilon)}{N_2(\varepsilon)} \approx \frac{\int_a^b c_{10}(x) dx}{\int_a^b c_{20}(x) dx} \tag{3.10}$$

with the right-hand side independent of $\varepsilon$ and with an error bound in the order $O(\varepsilon)$.

Since the quantity $l$ is taken to be the length between the two electrodes where the boundary conditions (3) are imposed, it would be much larger than the length of the channel. For example, if we take $l = 10$ (nm), which is four times the length $2.5$ (nm) used for $l$ in the estimation in (9), then the quantity $\varepsilon$ in (9) would be reduced by a factor of $1/4$. As a result, the factor of $9$ caused by the change of $\bar{\varepsilon}_r$ from $1$ to $80$ makes about $1.25\%$ difference in the approximation (10).

**When conclusion (8) might fail.** Recall that our conclusion, from (6) to (8), that the ratio does not depend significantly on the dielectric coefficient generally requires also condition (7), in addition to $\varepsilon$ being small. It is clearly possible that condition (7) does not hold; for example, in [3], when selectivity of $Na^+$ vs. $Ca^{2+}$ was examined for DEKA Na channel, a ***small*** amount (1 mM) of $CaCl_2$ was added to NaCl (100 mM) bulk solution so that Ca-concentration inside the channel is of order $O(10^{-3})$ relative to Na-concentration (see Fig. 3 in [3]). In this case, the conclusion (8) might fail. This is not surprise and an explanation for this possible failure goes as follows. Suppose, on the contrary, that the conclusion (8) still holds for this case. Since the numerator of the right-hand side of (8) is of order $O(1)$ due to the dimensionless scaling, the denominator would be order $O(10^{-3})$ which is comparable to $\varepsilon$; that is, the condition (7) does not hold. Going back to formula (6), the second term, due to its much smaller denominator, might not be small relative to the first term anymore. In this situation, the approximation (8) for the ratio fails in general; that is, the ratio would depend on the relative dielectric coefficient in more significant way. In fact, this is indeed the case as shown in Fig. 10 in [3]. This is consistent with the requirement of (7) in our result. Looking closer at Fig. 10 in [3], from Fig. 10 A, one sees that for $\varepsilon = 80$, the ratio is not large which indicates that (7) might hold (although not always since the numerator of the right-hand side of (8) might be small so that the denominator could still be small). This would imply our conclusion holds, that is, for $\varepsilon$ near $80$, the ratio does not depend on $\varepsilon$ significantly. This statement agrees well with the result shown in Fig. 10 B over the region $[40,80]$ for $\varepsilon$ which shows the slope of the ratio against $\varepsilon$ over this region is less than $(4-1.5)/(80-40) \sim 6\%$.

### 3.2   Ratio of fluxes of different ion species

For the same reason, the ratio of fluxes of two different ion species does not depend on the value of dielectric coefficient in a significant way. Indeed, if $J_1(\varepsilon)$ and $J_2(\varepsilon)$ are the fluxes of two ion species, then, from (3),

$$\frac{J_1(\varepsilon)}{J_2(\varepsilon)} = \frac{lC_0D_0J_{10} + \varepsilon lC_0D_0J_{11}(\varepsilon)}{lC_0D_0J_{20} + \varepsilon lC_0D_0J_{21}(\varepsilon)} = \frac{J_{10} + \varepsilon J_{11}(\varepsilon)}{J_{20} + \varepsilon J_{21}(\varepsilon)}$$

$$= \frac{J_{10}}{J_{20}} + \varepsilon \frac{J_{11}J_{20} - J_{10}J_{21}}{J_{20}^2} + O(\varepsilon^2). \tag{3.11}$$

Similar to condition (7),

$$\textit{if } |J_{20}| \textit{ is not small relative to } \varepsilon, \tag{3.12}$$

then

$$\frac{J_1(\varepsilon)}{J_2(\varepsilon)} \approx \frac{J_{10}}{J_{20}}$$

with the right-hand side independent of $\varepsilon$ and with an error bound in the order of $O(\varepsilon)$. Hence, when $\varepsilon$ is small, the ratio of the two fluxes does not depend on the value of dielectric coefficient significantly.

## 4   Conclusion

Some continuum theories of ion conduction allow reasonably general conclusions concerning physiological parameters and their dependence on underlying physical parameters. The PNP theory of open ionic channels can be analyzed by quite general mathematics to understand the dependence of selectivity on the relative dielectric coefficients. In this short paper, based on PNP type models that allow finite sizes of ions, we provide theoretical justifications when the ratio of the total numbers of two ion species does not depend on the relative dielectric coefficient in a significant way and when it might. A similar result is established for the ratio of fluxes of two ion species. In particular, the result explains that the selectivity does not depend on the relative dielectric coefficients in a significant way for ionic mixtures of $K^+Cl^-$ and $Na^+Cl^-$ but may depend on the relative dielectric coefficients in a significant way for ionic mixtures of $Na^+Cl^-$ and $Ca^{++}Cl_2$ which has been observed from extensive Monte Carlo simulations for DEKA Na channel in [3]. We believe that PNP theory of open ionic channels can provide an alternative way to understand ion channel properties.

## Acknowledgement

The authors thank Liwei Zhang for her numerical simulations in Figure 1. The authors thank the anonymous reviewers for their comments and suggestions that helped improve the manuscript.